\documentclass[12pt,preprint]{aastex}

\slugcomment{}

\shorttitle{NLTE red giant models}
\shortauthors{Short and Hauschildt}

\begin{document}


\title{Atmospheric Models of red giants with massive scale NLTE}


\author{C.I. Short}
\affil{Department of Astronomy \& Physics and Institute for Computational Astrophysics, Saint Mary's University,
    Halifax, NS B3H 3C3}
\email{ishort@ap.stmarys.ca}

\author{P.H. Hauschildt}
\affil{Hamburger Sternwarte, Gojenbergsweg 112, 21029 Hamburg, Germany}
\email{phauschildt@hs.uni-hamburg.de}



\begin{abstract}

  We present plane-parallel and spherical LTE and NLTE atmospheric models 
of a variety of stellar parameters
of the
red giant star \objectname{Arcturus} ($\alpha$ Boo, \objectname{HD124897}, \objectname{HR5340})
and study their ability to fit the measured absolute
flux distribution.
Our NLTE models include tens of thousands of the strongest lines in 
NLTE, and we investigate separately the effects of treating 
the light metals and the Fe group elements Fe and Ti in NLTE.  
We find that the NLTE effects of Fe group 
elements on the model structure and $F_\lambda$ distribution are much more 
important than the NLTE effects of all the light metals combined,
and serve to substantially increases the violet and near UV
$F_\lambda$ level as a result of NLTE Fe over-ionization.  
Both the LTE and NLTE models predict significantly more flux in the 
blue and UV bands than is observed.
We find that within the moderately metal-poor metallicity range,
the effect of NLTE on the overall UV flux level {\it decreases} with
decreasing metallicity.
These results suggest that there may {\it still} be
important UV opacity missing from the models. 
We find that models
of {\it solar metallicity} giants of similar spectral type to Arcturus fit well 
the
observed flux distributions of those stars from the red to the near UV band.
This suggests that the blue and near UV flux discrepancy is metallicity dependent,
increasing with decreasing metallicity.

\end{abstract}




\section{Introduction}

  Because red giant stars are relatively bright and more numerous than other luminous stars, they
serve as chemical abundance tracers for remote locations, both in the various
structures of our own galaxy and in other galaxies.  In this capacity 
red giant stars have taken on additional importance recently as a result
of the discovery of a relic population in the galactic halo that
are known variously as Very Metal Poor (VMP), Ultra Metal Poor (UMP), or
Extremely Metal Poor (XMP) stars \citep{cohen1}.  However,
the derivation of chemical abundances from flux distributions and spectra is 
highly model dependent.  Therefore,
the physical realism of atmospheric models and spectrum synthesis calculations plays an important role in determining
the acuity with which we can distinguish abundance patterns throughout space.
The most widely used grids of models for red giant chemical abundance analyses
to date have been the ATLAS series of models \citep{kurucz94}
and the MARCS (most recently NMARCS) series of models
\citep{plez_bn92}.
Both sets of models make use of
the assumption of Local Thermodynamic Equilibrium (LTE), and the former makes use of the assumption
of plane-parallel geometry.
The latest LTE models have
been successful at fitting the global solar flux distribution, but, it is
unclear whether the expanded line lists provide a satisfactory fit to the blue and UV
bands of red giant stars.
\citet{bessell_cp98} undertook a critical evaluation of how well these 
grids of atmospheric models fit empirical relations between stellar color
and the fundamental stellar parameters of effective temperature ($T_{\rm eff}$)
and bolometric magnitude ($M_{\rm bol}$).
For giants of spectral types G to M, none of
the models can fit the relations between $T_{\rm eff}$ and those colors that
are derived from the blue or violet regions of the spectrum, such as Johnson
$B-V$.
They postulate that incomplete or erroneous opacity in the blue-violet region
is responsible for the discrepancy.
Similarly, \citet{short_l96} found that ATLAS9 models over-predict the emergent flux by a factor of two,
and systematically over-predict the strength of all spectral lines, in the violet spectral
region ($\lambda \le 4200$ \AA) of red giants like Arcturus, unless the {\it continuous} opacity is
approximately doubled, {\it ad hoc}.

\paragraph{}

Red giant atmospheres are relatively translucent over large path lengths and non-locally determined 
radiative rates often dominate collisional rates for many transitions, causing departures from local thermodynamic
equilibrium (LTE).  Previous investigations of exploratory scope \citep{gratton_ceg99}, \citep{thevenin_i99} have found that   
abundances of red giant stars that are derived from {\it non}-LTE line profile calculations can differ significantly from those
derived from LTE calculations.  Furthermore, the discrepancies become larger as the metallicity is reduced to
significantly sub-solar values.   
Therefore, {\it both} the atmospheric structure and the emergent spectrum
should be calculated in the more general statistical equilibrium (SE), in
which a set of coupled equations is solved for the rate at which every
energy level of every ionization stage of every species is populated
and de-populated by various collisional and radiative processes.
Given the importance to structure formation studies of chemical abundance analysis of the 
recently discovered
XMP stars in the galactic halo, more realistic modeling of 
metal poor red giants is particularly important.  

\paragraph{}

   As a result of computational constraints, previous investigations of NLTE effects in red giant stars have treated a few 
chemical species in SE while treating the bulk of the strongest lines in LTE. 
Recently, \citet{short_hb99} have greatly expanded the NLTE SE treatment in the general-purpose 
atmospheric modeling and spectrum
synthesis code {\tt PHOENIX} \citep{hauschildt_b99} so that over 100\,000 spectral lines, 
including all of the 
strongest lines and many of the weaker lines that blanket the UV band, 
are now treatable in self-consistent NLTE.  Even this is a small fraction of
the millions of spectral lines that collectively control the emergent flux, but it is a
significant step forward in improving the realism of the models.  The purpose of this study is to
calculate theoretical models and synthetic spectra for the mildly metal-poor standard red giant 
star Arcturus
($\alpha$ Boo, HD124897, HR5340) to assess the affect
that large scale NLTE line blanketing has on the theoretical model structure and the synthetic
spectrum.  The modeling of a mildly metal-poor standard red giant like Arcturus
with large scale SE and spherical geometry is the first step in a larger
effort to model the XMP stars with unprecedented realism. 
In Section \ref{sec_model} we describe the computational 
modeling;
in Section \ref{sec_stelpar} we discuss the stellar parameters; in 
Section \ref{sec_reslt} we present our results for both the Sun and
Arcturus, and re-iterate out main conclusions in Section \ref{sec_con}.

\section{Modeling \label{sec_model}}

{\tt PHOENIX} makes use of a fast and accurate Operator Splitting/Accelerated Lambda Iteration
(OS/ALI) scheme to solve self-consistently the radiative
transfer equation and the NLTE SE rate equations for many species and overlapping transitions
\citep{hauschildt_b99}.
Recently \citet{short_hb99} have greatly
increased the number of species and ionization stages treated in
SE by {\tt PHOENIX} so that at least the lowest two stages of 24 elements,
including the lowest six ionization stages of the 19 most
important elements, are now treated in NLTE.  \citet{short_hb99} contains details
of the sources of atomic data and the formulae for various atomic processes.
Table 1 shows which species have been treated in NLTE in the modeling presented here, and how many $E$ levels and $b-b$ (bound-bound) transitions are included in 
SE for each species.  For the species treated in NLTE, only transitions of $\log gf$ value
greater than -3 (designated primary transitions) are included in the SE equations.  
All other transitions of that species (designated secondary transitions) are treated 
in LTE.  
We have only included in the NLTE treatment those ionization stages that are non-negligibly 
populated at some depth in the Sun's atmosphere.  As a result, we only include the
first one or two ionization stages for most elements.  

\paragraph{}

   Table 2 shows the six levels of realism with which we calculate our models
and synthetic spectra.  The six levels incorporate three degrees of realism
in the treatment of the equilibrium state of the gas and the radiation field,
and two degrees of realism in the geometry of the atmosphere.  Unless otherwise
noted, the realism of a synthetic spectrum calculation is always consistent with
that of the input model used.  
We have investigated the behavior of NLTE models with two levels
of realism: 1) NLTE treatment for H, He, and important 
light metals up to, but {\it not} including, the Fe group elements (designated NLTE$_{\rm Light}$ models), 
and 2) the same as the NLTE$_{\rm Light}$ models except that the Fe group elements Fe and Ti 
are also included in the NLTE treatment (designated NLTE$_{\rm Fe}$ models).  We investigate
these two levels of realism because the Fe group elements play a special role
in the atmospheres and spectra of late-type stars \citep{thevenin_i99}; because of their spectacularly rich term
structure a neutral or low ionization stage Fe group element contributes between 
one and two orders of magnitude more lines to the spectrum than the corresponding
stage of any lighter element.  
Finally, we note that all of the models in
Table 2 also include many tens of millions additional lines from many atoms,
ions and diatomic molecules in the approximation of LTE.   
\citet{hauschildt_afba99} have calculated cool super-giant models with spherical geometry and found that
the broad-band emergent flux is often under-predicted by as much as 50\% by plane parallel models 
due to the effect of sphericity on line blanketing opacity.  Therefore, we include an
investigation of geometry here.

\section{Stellar parameters \label{sec_stelpar}}
 
%


  The most recent determination of the fundamental parameters of Arcturus
is that of \citet{griffin_l99}.  They compiled observations
of the absolute flux distribution, $F_\lambda$, from the UV to the IR 
from several sources and found a total integrated flux of 
$4.95\times 10^{-5}$ ergs cm$^{-2}$ s$^{-1}$.  This they combined with a recent 
determination of the angular diameter, $\theta$, by \citet{quirren} of
$21.0\pm 0.2$ mas to derive a $T_{\rm eff}$
value of $4291\pm 27$K.  
They also found the closest match to 
their composite $F_\lambda$ distribution among a grid of calculated 
$F_\lambda$ distributions that were generated with the atmospheric models of
\citet{kurucz92b} and found best fit values for $T_{\rm eff}$, $\log g$,
and $[{\rm A}/{\rm H}]$ of $4291.9\pm 0.7$K, $1.94\pm 0.05$, and $-0.68\pm 0.02$.
The close agreement
between the $T_{\rm eff}$ values determined from two independent methods
is reassuring, and both the $T_{\rm eff}$ value and the $\log g$ value are
within the range of most previous determinations.  Their value of 
$[{\rm A}/{\rm H}]$ is near the lower end of the range of previously determined
values.  

\paragraph{}

A recent thorough determination of the
parameters using another technique, LTE spectral fitting of the 
profiles of many lines throughout the visible band, by \citet{pdk93},
yielded the values $T_{\rm eff}=4300\pm 30$K,
$\log g =1.5\pm 0.15$, and a value $[{\rm A}/{\rm H}]=-0.5\pm 0.1$ for most elements,
with $\alpha$-process elements twice as abundant.  The $T_{\rm eff}$ value of
the latter study is in very close agreement with both values derived by 
\citet{griffin_l99}, has a $\log g$ value near the lower end of the generally 
accepted range for Arcturus, is more metal rich by $0.2$ dex, and has a 
non-solar abundance pattern.  

\paragraph{}

  \citet{decin1}, \citet{decin2} and \citet{decin4} have fit synthetic spectra based on plane-parallel and spherical LTE
atmospheric models to the near IR band (2.38 to 4.08 $\mu$m) as observed by the Short-Wavelength Spectrometers
of the Infrared Space Observatory (ISO-SWS) for a wide range of late-type giant stars, including Arcturus.
Their modeling was part of a program to simultaneously investigate the calibration of ISO-SWS and the
realism of atmospheric models.  Therefore, they performed a thorough analysis of the dependence of
the strength of individual molecular features throughout the 2.38 to 4.08 $\mu$m band on the value of
individual stellar parameters.  \citet{decin4} find the following best fit parameters: T$_{\rm eff}=4320\pm 140$ K,
$\log g=1.50\pm 0.15$, $[{\rm A}/{\rm H}]=-0.5\pm 0.20$, and $\xi_{\rm t}=1.7\pm 0.5$ km s$^{-1}$.  They also
found that C/Fe, N/Fe, and O/Fe are enhanced by 0.1, 0.4, and 0.4, respectively, with respect to
the solar ratios.  The values of T$_{\rm eff}$ and $[{\rm A}/{\rm H}]$ are in agreement, to within the
uncertainties, of both \citet{griffin_l99} and \citet{pdk93}, and the $\log g$ value agrees with that of \citet{pdk93}.

\paragraph{}

Because Arcturus is a thick disk population (mildly Population II)
star, one may expect a non-solar abundance distribution such as that
reported by \citet{pdk93} and \citet{decin4}.  Indeed, the
possibility of detecting such a pattern by way of accurate modeling and
spectrum synthesis is important in the broader context of galactic
chemical evolution, and is the motivation for the improved modeling
presented here.
However, our purpose in this initial assessment of the massively NLTE models is to
investigate the extent to which the incorporation of
NLTE effects for most of the strong atomic lines affects the
calculated equilibrium model structure and emergent $F_\lambda$ distribution.
To that end we adopt a canonical model with a scaled solar abundance distribution, designated
T43G20M07, with the parameters of \citet{griffin_l99},
rounded to the nearest cardinal values typically used in model atmosphere grids:
$T_{\rm eff}=4300$K, $\log g=2.0$, and $[{\rm A}/{\rm H}]=-0.7$.  For exploratory
purposes we also study four models at adjacent points that span the
range of values in $T_{\rm eff}$,
$\log g$, and $[{\rm A}/{\rm H}]$ that have been found by various
investigators.  Table 3 displays the model designations and parameters
for this small grid.
T43G15M07, T43G20M04, T43G15M04, and T42G20M07.

\paragraph{}
Unique specification of a spherical model requires a parameter in addition
to $T_{\rm eff}$, $\log g$, and $[{\rm A}/{\rm H}]$.  We set the radius at a standard
optical depth of unity $R(\tau_{12000}=1)$ equal to $23$R$_\odot$ in
keeping with the value that \citet{griffin_l99} found from fitting the overall
flux distribution and the measured angular diameter.
The specification of
$\log g$ and $R$ sets the mass, $M$, by way of the formula for surface gravity,
$g=GM/R^2$, to 2M$_\odot$.  The specification of $T_{\rm eff}$ and $R$ sets
the bolometric luminosity, $L_{\rm Bol}$, to $4\pi R^2\sigma T_{\rm eff}^4$ (by
the definition of $T_{\rm eff}$), which is 160 L$_\odot$ ($M=-0.7$).  By comparison,
\citet{decin4} used the same relations with their best-fit stellar parameters to
find: $R(\tau_{\rm Continuum}=1)=25.06\pm 1.56$R$_\odot$, $M=0.75\pm 0.27$M$_\odot$, and
$L=197\pm 36$L$_\odot$.  Their lower value of $M$ is consistent with the lower
value of $\log g$ that they adopted.
We adopt a depth-constant micro-turbulent velocity dispersion, $\xi_{\rm T}$,
 of $2.0$ km s$^{-1}$.

\section{Results \label{sec_reslt}}

\subsection{Model structure}

\begin{figure}
\plotone{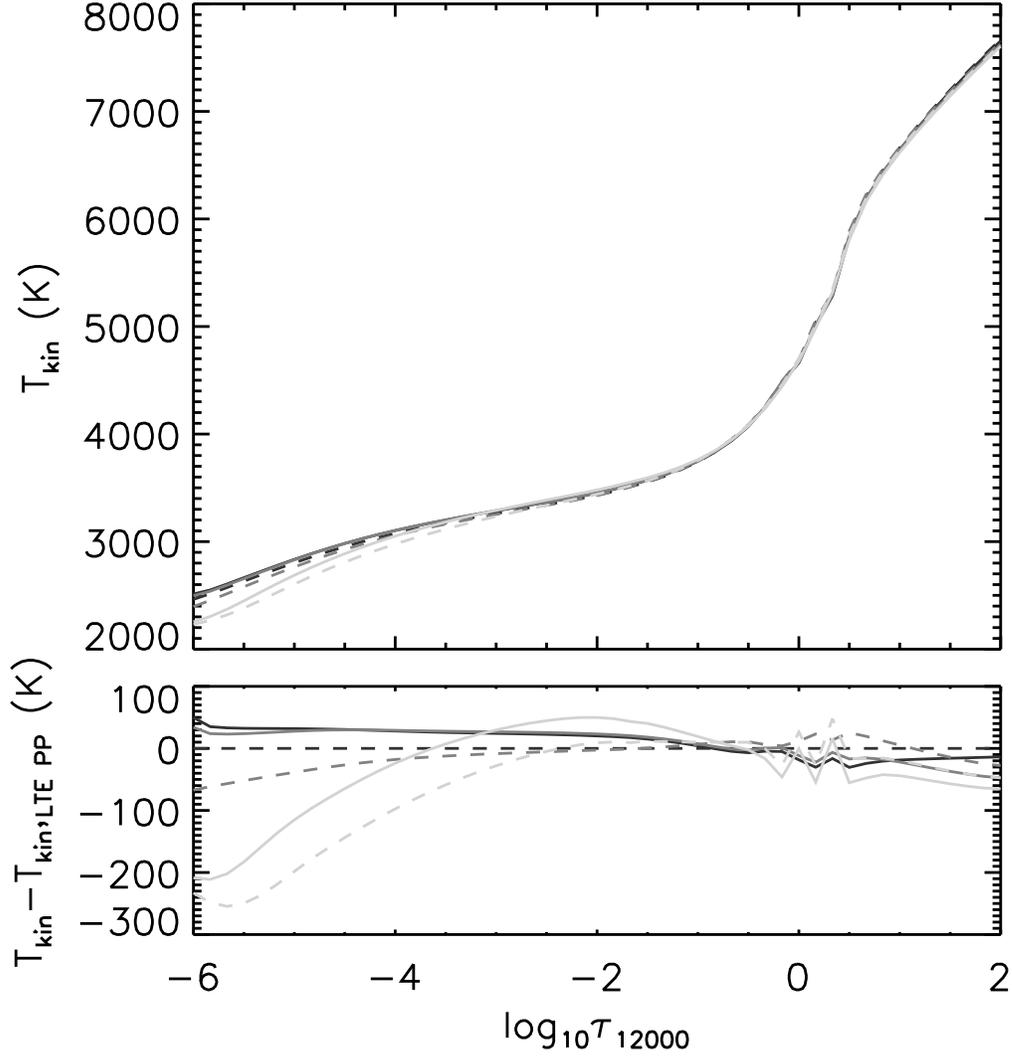}
\caption{Temperature ($T_{\rm kin}$) structure of theoretical atmospheric models of Arcturus computed with: 
LTE (dark lines), light metals in NLTE (NLTE$_{\rm Light}$) (medium lines),
light metals and Fe and Ti in NLTE (NLTE$_{\rm Fe}$) (light lines).
For each degree of NLTE, the models were computed with spherical geometry 
(solid lines) and plane-parallel ($PP$) geometry (dashed lines).
Upper panel: Absolute $T_{\rm kin}$, lower panel: $T_{\rm kin}$ relative to
that of the LTE $PP$ model. 
\label{stars_mods}}
\end{figure}

Fig. \ref{stars_mods} shows the computed kinetic temperature ($T_{\rm kin}$) structure 
for the canonical Arcturus models (T43G20M07) with the different levels of physical
realism.  
We use the LTE $PP$ model as a reference model for the comparison in the lower panel.  
The atmosphere is sufficiently
extended compared to the radius of the star that the effects of sphericity are
significant.  Especially interesting is the dependence of the
sphericity effect on the extent to which the model deviates from LTE. 
Comparing the dashed lines in Fig. \ref{stars_mods} we see a trend of
increasing surface cooling with increasing NLTE effects among the $PP$
models. 
The $PP$ NLTE$_{\rm Light}$ and NLTE$_{\rm Fe}$ models are cooler than the $PP$ 
LTE model by more 
than $50$ and $250$K at the top of the atmosphere, respectively.
By contrast, comparing the solid lines we see that the spherical NLTE$_{\rm Light}$ 
model deviates negligibly 
($\le 10$K) from the spherical LTE model.  As a result, with $PP$ modeling one
would overestimate the effect of NLTE among light metals on the
model structure.  Both the spherical LTE
and spherical NLTE$_{\rm Fe}$ models are about $50$K warmer near
the top of the atmosphere than their
LTE counterparts, so among the spherical models the NLTE$_{\rm Fe}$ is
still about $250$K cooler than the LTE model.   In every case, 
spherical models are always warmer throughout the outer atmosphere than $PP$
models.
In their study of sphericity effects in line blanketed giants and
super-giants, \citet{hauschildt_afba99} also found that spherical models
were warmer than corresponding $PP$ models (see their Fig. 8).  
 
\subsection{Absolute flux distribution}

\begin{figure}
\plotone{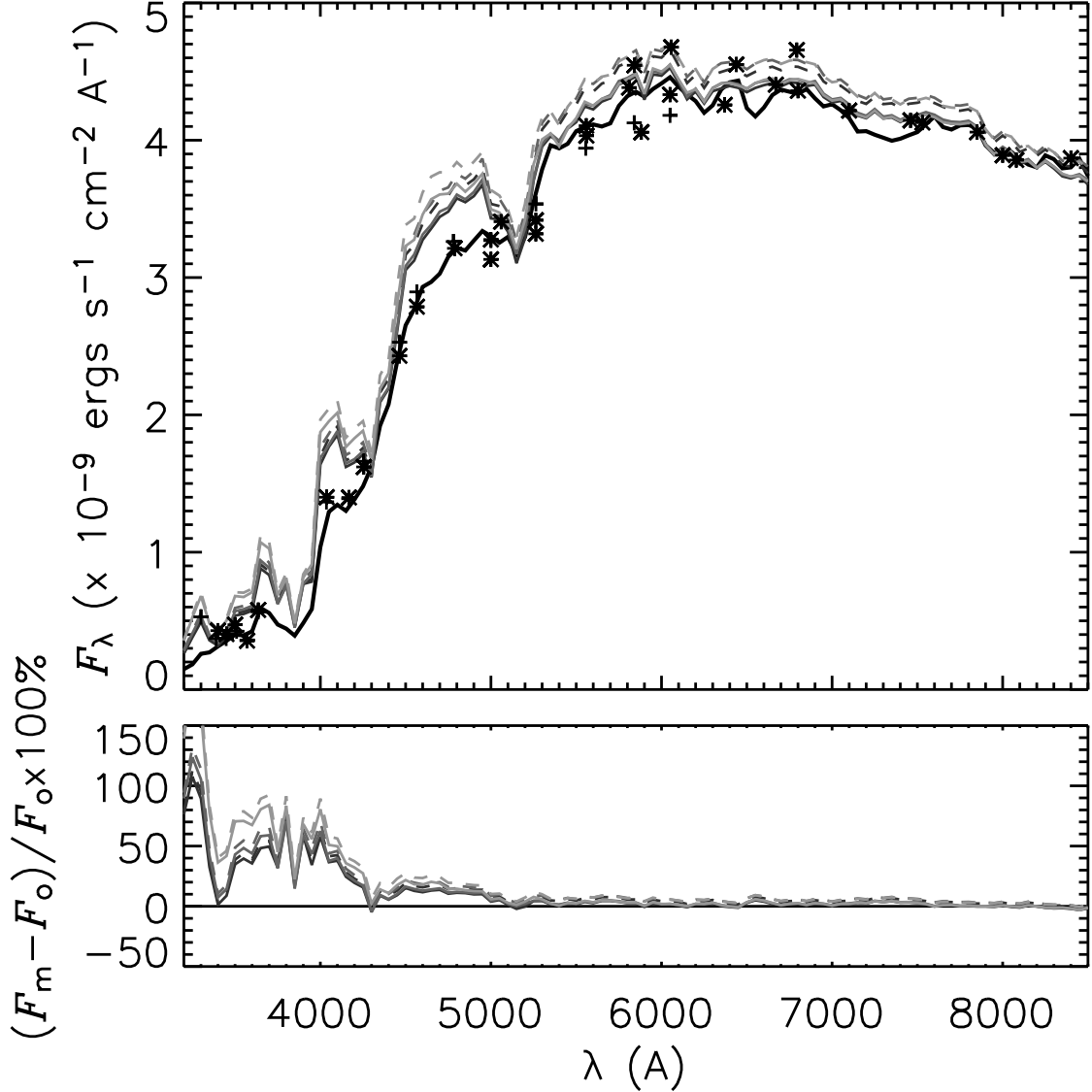}
\caption{Comparison of the observed (thick black line) and computed 
$F_\lambda(\lambda)$ distribution.  The asterisks are intermediate band spectrophotometry 
from \citet{breg}.  Computed distributions are shown for
models calculated in $PP$ (dashed lines) and spherical (solid lines) geometry
for the LTE (dark thin line), NLTE$_{\rm Light}$ (medium line), and NLTE$_{\rm Fe}$ (light line) models.
Upper panel: Absolute $F_\lambda$, lower panel: the difference between the model ($F_{\rm m}$) and observed ($F_{\rm o}$) $F_\lambda$ distributions, as a
percentage of observed $F_\lambda$. \label{stars_flxall}} 
\end{figure}

 Fig. \ref{stars_flxall} shows the overall $F_\lambda$ distribution of
the canonical model (T43G20M07) calculated with all levels of physical
realism, diluted to the distance of
Arcturus using the measured angular diameter of 21.0 mas \citep{quirren}.  
The observed $F_\lambda$ distributions are that of \citet{pulk5} and
\citet{breg}.
The computed distributions have a spectral resolution, $R$, of 50\,000 and we have
convolved them with a Gaussian of FWHM equal to 
50\AA~ to match the sampling of the 
\citet{pulk5} data.   We also show the 
difference between the various computed $F_\lambda$ distributions and the
observed distribution as a percentage of the observed distribution.  
All computed distributions are in general agreement with the over-all
flux level in the red and yellow bands of the 
\citet{pulk5} short
wavelength data-set.  However, all over-predict the flux by as much as 
150\% with decreasing $\lambda$ into the blue and UV bands.  The 
spherical models have less flux than the $PP$ models by as much as 10\%
throughout the blue and UV, thereby providing a fit that is marginally closer, 
but still poor.  

%
%
%

\paragraph{}

The LTE and NLTE$_{\rm Fe}$ models differ negligibly from each other for 
$\lambda > 5000$\AA, but for shorter $\lambda$ values the NLTE$_{\rm Fe}$ model 
becomes increasingly brighter than the LTE model.  As a result, with these stellar
parameters the LTE model provides a marginally closer fit to the observed $F_\lambda$ level
than the more realistic NLTE$_{\rm Fe}$ model.  

\paragraph{}

\begin{figure}
\plotone{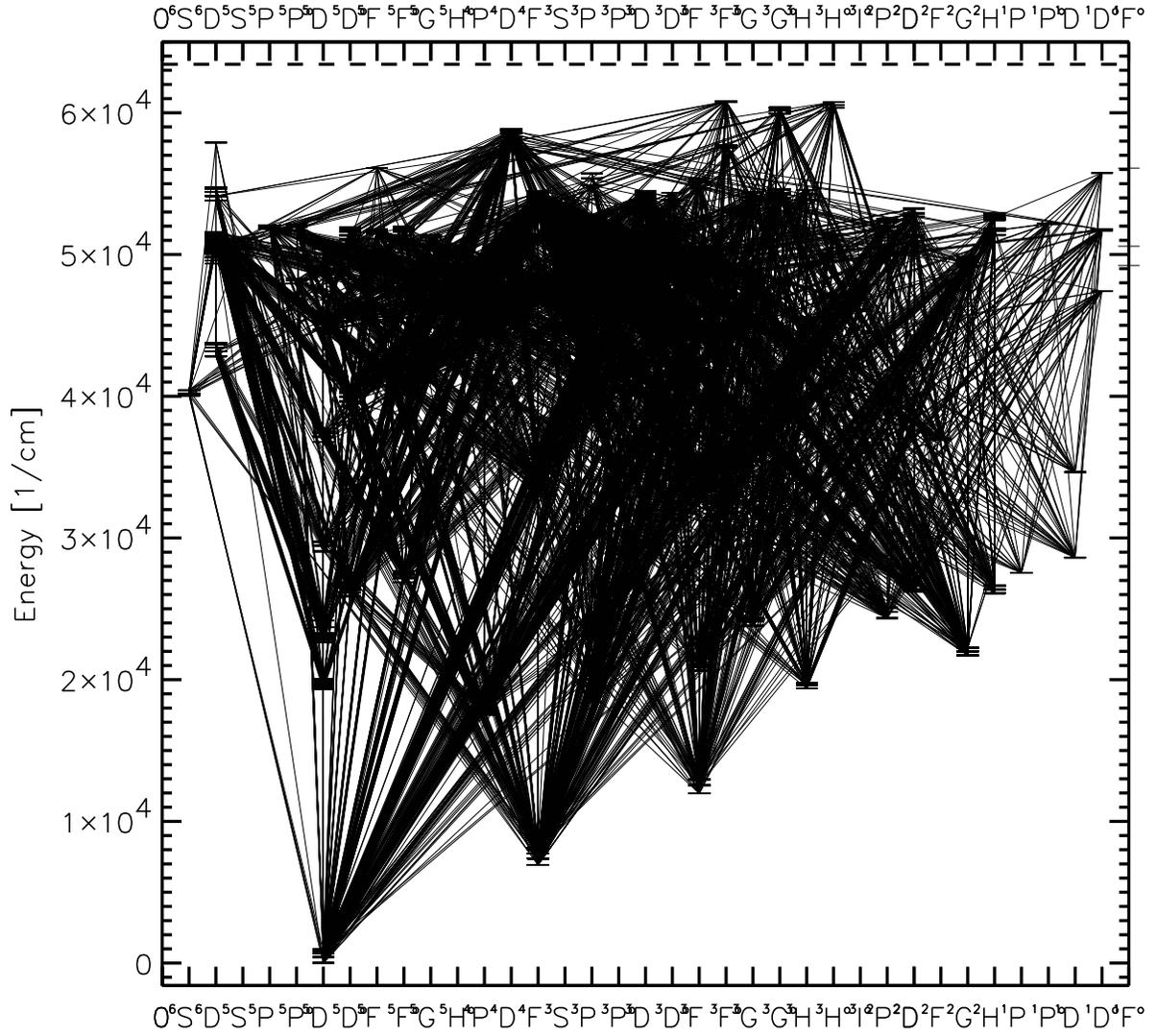}
\caption{Grotrian diagram of the model \ion{Fe}{1} atom used in 
our NLTE calculations.
\label{fe_grot}} 
\end{figure}

\begin{figure}
\plotone{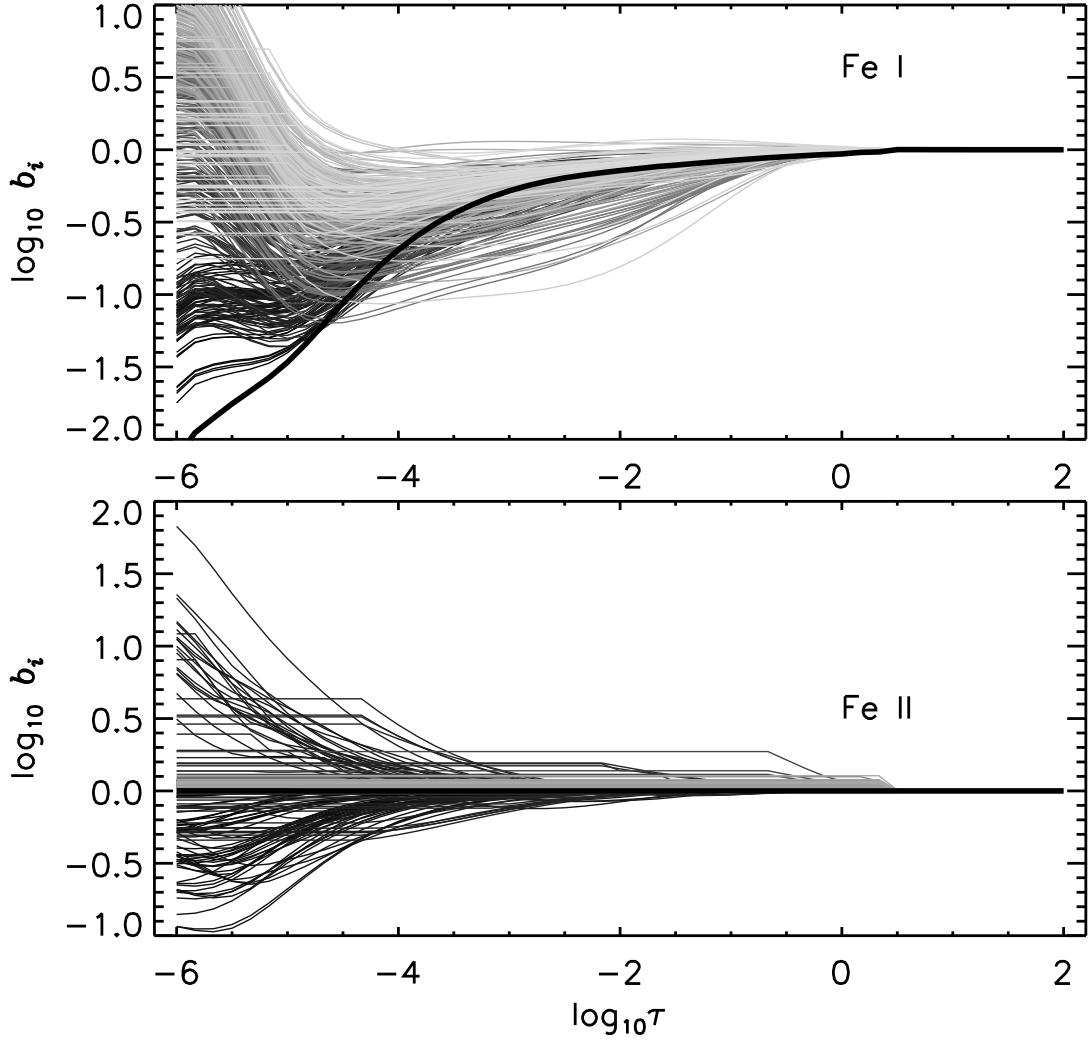}
\caption{NLTE departure coefficients for \ion{Fe}{1} and \ion{}{2} in the
NLTE$_{\rm Fe}$ model of Arcturus.  The 
ground state coefficient is shown with a thick black line.  The lighter
the color of the line the higher the energy, $E$, of the level.
\label{stars_bi}} 
\end{figure}

  The reason for the increased $F_\lambda$ level in the UV with the
NLTE$_{\rm Fe}$ model can be seen in Figs. \ref{fe_grot} and \ref{stars_bi}.
Because of its rich term structure, Fe contributes a 
significant fraction of the total line opacity, particularly in the UV band.  
Although the \ion{Fe}{1} $b_{\rm i}$ values are larger than unity for many of the
higher $E$ levels in the outermost part of the atmosphere, the ground
state and all the lower $E$ levels are under-populated with respect
to LTE.  This is due to an over-ionization of Fe
with respect to LTE that depletes the \ion{Fe}{1} population throughout
the atmosphere.  As a result, the veil of Fe opacity in the UV band is weakened
which allows more flux to escape in the UV. 

\paragraph{}
We re-emphasize that our calculations are carried
out using the the expanded line list of \citet{kurucz92a}, which was found to provide a 
close fit the the UV flux level of the Sun.  Apparently, the ability of the
expanded line list to address the solar UV flux problem does not extend to
all red giants.  
Given the magnitude of the differences in the $T_{\rm kin}$
structure of the outermost depths between the LTE and NLTE models, seen
in Fig. \ref{stars_mods}, we have also performed an internally inconsistent
calculation of the $F_\lambda$ distribution with the NLTE$_{\rm Fe}$ set-up using the 
LTE model as input.  This $F_\lambda$ distribution lies very close to that of the self-
consistent NLTE$_{\rm Fe}$ calculation, which indicates that the direct effects
of NLTE radiative transfer and statistical equilibrium are a much larger
influence on the spectrum formation than the effect of 
NLTE atmospheric structure.  Given the agreement between the LTE and
NLTE models at deeper layers around $\tau_{12000}=1$, the relative
unimportance of the atmospheric structure in determining the emergent
$F_\lambda$ is not surprising.

\paragraph{}
  
Excess model flux in the UV suggests that the $T_{\rm eff}$
value of the model (4300K) may be too high, although the
range of recent determination from a variety of methods is constrained to
within 50K.  
Furthermore, previous $\log g$ and $[{\rm A}/{\rm H}]$ determinations have ranged from our values
 of 2.0 and -0.7 to 1.5 and -0.4.  
To test whether NLTE models provide a better fit with other 
values within the error range of the stellar parameters, we
converged a small grid of four spherical LTE and NLTE$_{\rm Fe}$ models in the 
$\log g - [{\rm A}/{\rm H}]$
plane, all with a $T_{\rm eff}$ value of 4300K.  The grid is displayed in 
Table 3.  
Furthermore, we converged spherical LTE and NLTE$_{\rm Fe}$ models with a 
$T_{\rm eff}$ value of $4200$K and our canonical $\log g$ and $[{\rm A}/{\rm H}]$ 
values of 2.0 and -0.7.  The latter model is well outside the error limit of recent
$T_{\rm eff}$ determinations, but we include it to demonstrate by how much 
$T_{\rm eff}$ must deviate from the canonical value to force a fit to the UV
flux level.

\paragraph{} 

\begin{figure}
\plotone{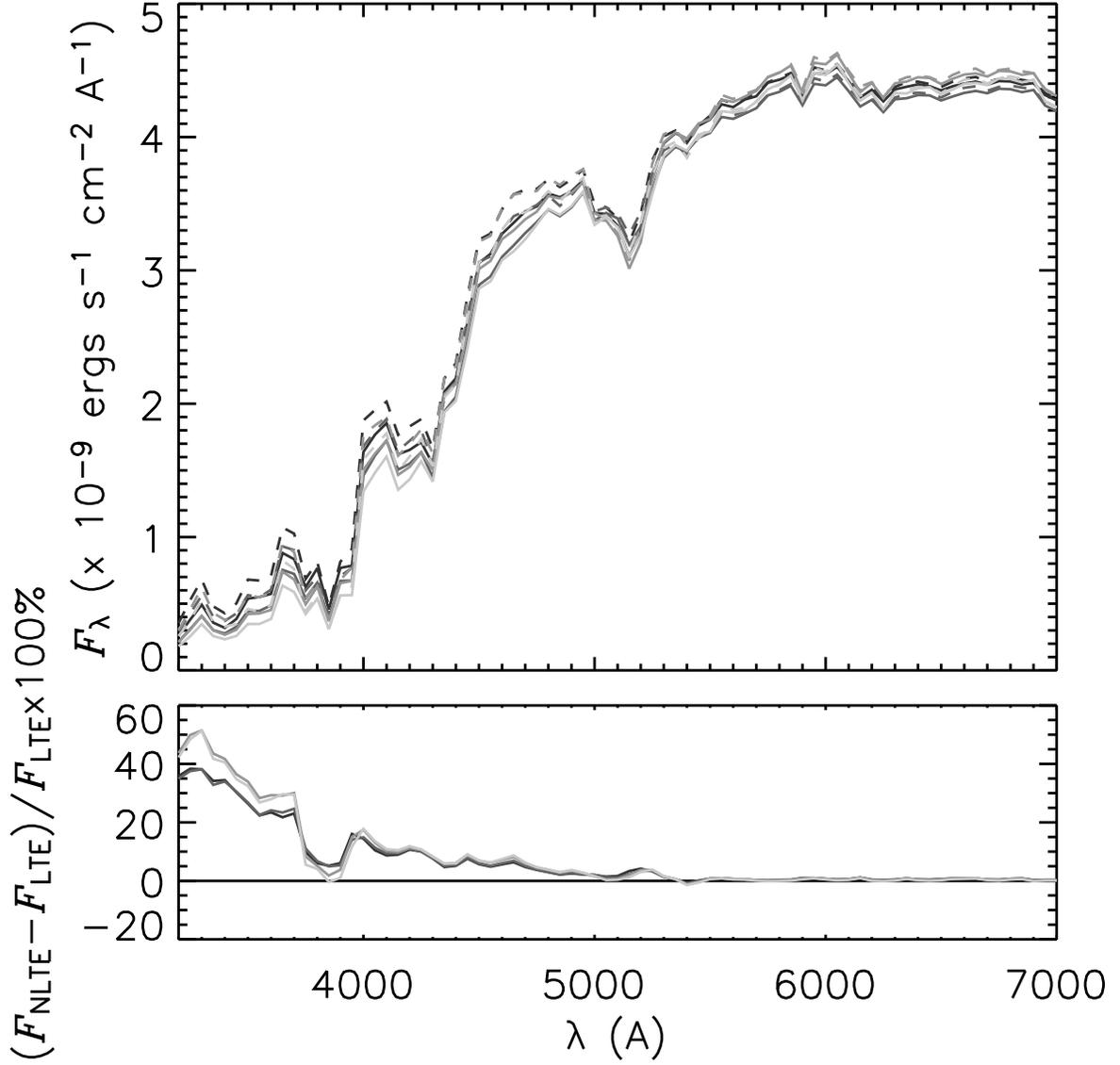}
\caption{Comparison among computed  
$F_\lambda(\lambda)$ distributions for the spherical LTE and NLTE$_{\rm Fe}$ models
throughout the parameter grid.  In order of decreasing line darkness,
T43G20M07 (canonical parameters), T43G15M07, T43G20M04, and T43G15M04.
Solid lines: LTE models; dashed lines: NLTE$_{\rm Fe}$ models.
Upper panel: Absolute $F_\lambda$ distributions; lower panel:
difference between NLTE and LTE models for each set of stellar parameters.  
\label{stars_flxmods}}
\end{figure}

Fig. \ref{stars_flxmods} shows the comparison among synthetic
$F_\lambda$ distributions of spherical LTE and NLTE$_{\rm Fe}$ models
in the $\log g - [{\rm A}/{\rm H}]$ grid.  All models 
provide $F_\lambda$ distributions that agree relatively closely for
$\lambda > 5000$ \AA, and become increasingly distinct from each other as
$\lambda$ decreases.  Both the decrease in $\log g$ and
the increase in $[{\rm A}/{\rm H}]$ serve to decrease the predicted
UV band $F_\lambda$ level.  The latter effect is expected due to the
increase in line blanketing with decreasing $\lambda$.  
There is an approximate ``degeneracy'' in 
$\log g$ and $[{\rm A}/{\rm H}]$
in that the T43G15M07 and T43G20M04 models give rise to UV $F_\lambda$
levels that are barely distinguishable.  Our calculations allow us to
assess the dependency of the magnitude of NLTE effects on the stellar
parameters.  The extent of the NLTE UV brightening due to Fe over-ionization
is sensitive to $[{\rm A}/{\rm H}]$, but not to $\log g$.  The
amount of NLTE UV brightening is as much as $10\%$ greater in the models
of $[{\rm A}/{\rm H}]=0.4$ than it is in models of 
$[{\rm A}/{\rm H}]=0.7$.  
Therefore, we find, for this metallicity range,
that the effect of NLTE on the overall UV flux level {\it decreases} with
decreasing metallicity. 

\paragraph{}

\begin{figure}
\plotone{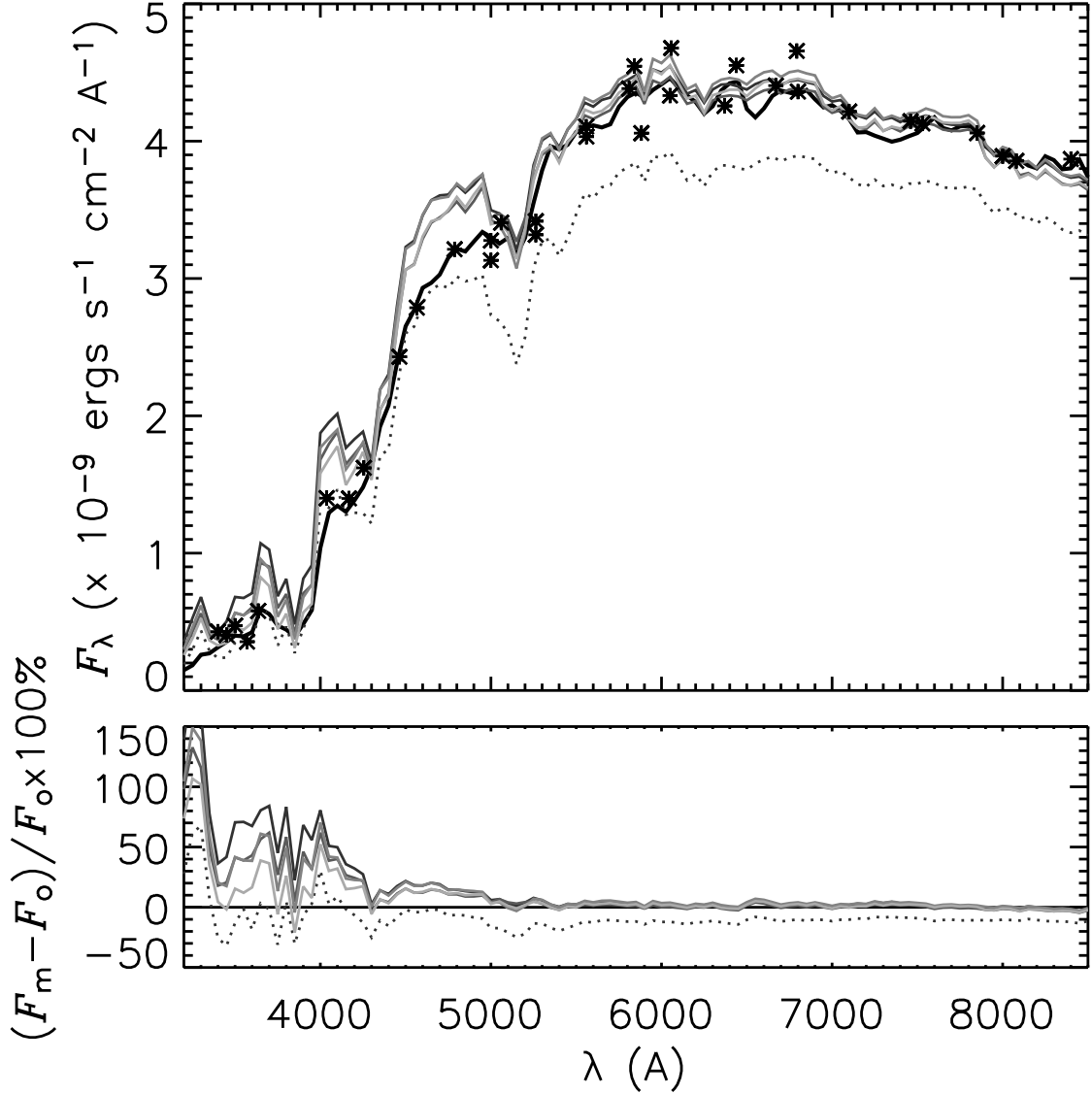}
\caption{Comparison of observed and computed 
$F_\lambda(\lambda)$ distributions for models throughout the 
parameter grid.  Observed spectrum: thick black line.  Theoretical distributions are 
shown for the spherical NLTE$_{\rm Fe}$ models, in order of decreasing line darkness,
T43G20M07 (canonical parameters), T43G15M07, T43G20M04, and T43G15M04; dotted 
line: T42G20M07.  
Upper panel: Absolute $F_\lambda$, lower panel: the difference between the model ($F_{\rm m}$) and observed ($F_{\rm o}$) $F_\lambda$ distributions, as a
percentage of observed $F_\lambda$.  
\label{stars_flxall2}}
\end{figure}

Fig. \ref{stars_flxall2} shows the comparison of synthetic
$F_\lambda$ distributions of spherical NLTE$_{\rm Fe}$ models in the
$\log g - [{\rm A}/{\rm H}]$ grid and of the $T_{\rm eff} = 4200$K 
model to the observed spectrum.  
All models provide approximately the same quality of fit to 
the observed $F_\lambda$ distribution throughout the yellow and red bands.
Even the model that is faintest in the UV, T43G15M04, is as much as $100\%$ 
too bright between 3000 and 3500 \AA.  
Therefore, throughout the $\log g - [{\rm A}/{\rm H}]$ 
range in which Arcturus is expected to lie, no model of $T_{\rm eff} = 4300$K
provides even an approximately close match to the observed blue and UV brightness.  
The $4200$K model provides a closer match to the overall observed UV $F_\lambda$ 
level, but is systematically too faint throughout the visible band where
$F_\lambda(\lambda)$ peaks and the Rayleigh-Jeans tail begins.  We conclude that the UV discrepancy is not due to
incorrect values of the stellar parameters.

 
\paragraph{}
 
   It should be noted that, generally, for late-type stars, $F_\lambda$ in 
the mid to far UV bands exceeds the values predicted by models in radiative 
and convective equilibrium because in these bands the flux arises from layers
above the temperature minimum region where the temperature distribution with
depth is inverted due to the presence of a chromosphere.  However, this
extra flux due to non-radiative heating of the outer atmosphere typically
arises only in the $\lambda < 2500$\AA~ region \citep{morossi_fmkb93}.  
Therefore, the presence of 
a chromosphere is not likely to be the cause of the blue and near UV band 
discrepancy found in K2\,{\sc III} stars.

\subsection{Comparison to other late-type giants}

\begin{figure}
\plotone{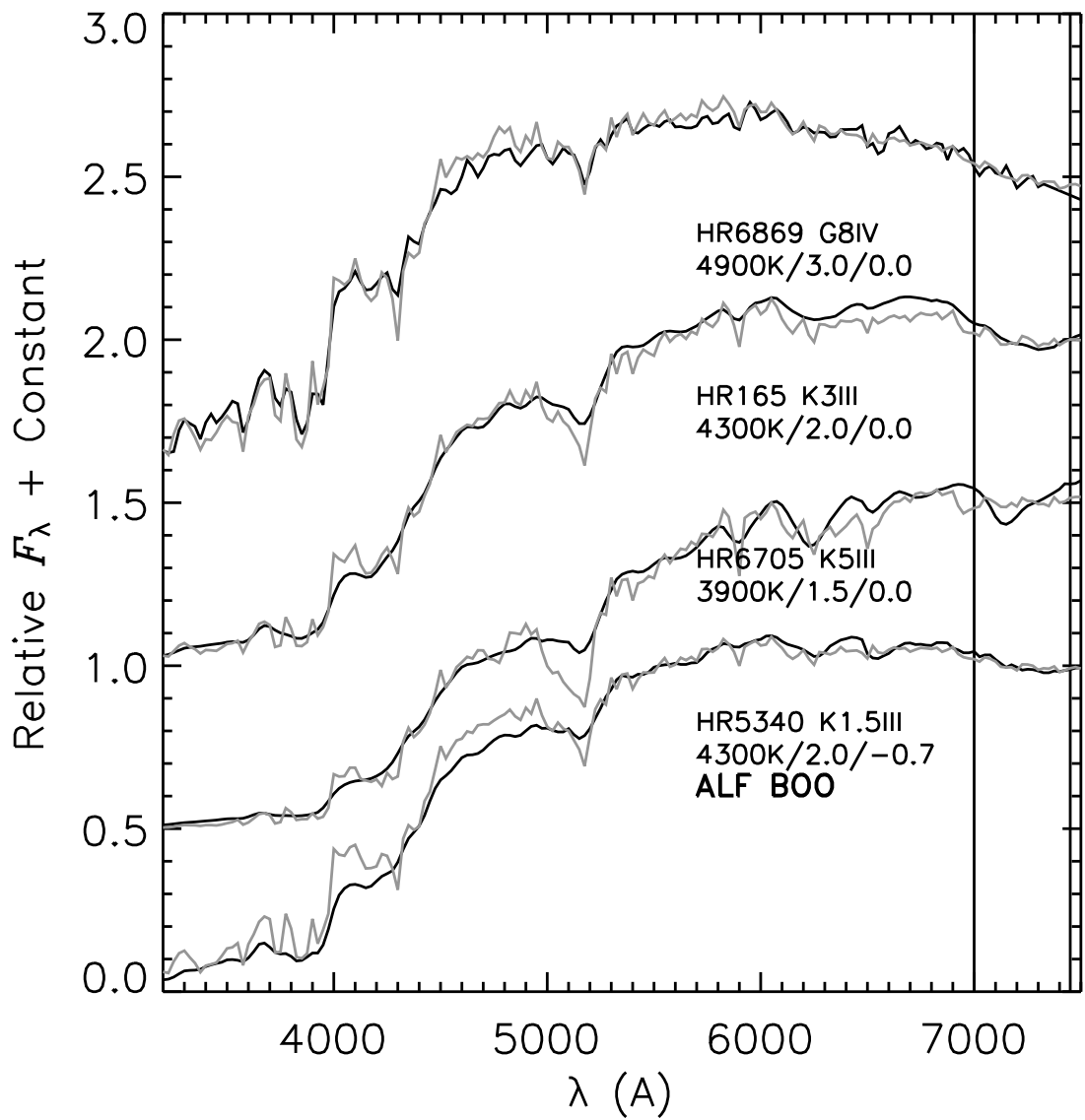}
\caption{Comparison of observed and $F_\lambda(\lambda)$ distributions to
theoretical distributions computed with spherical LTE models for a sample
of late-type
giant stars with reliable stellar parameters from \citet{morossi_fmkb93} (details in
Table 5).  Spectrophotometric data for all stars are from \citet{pulk5}
Observed and computed spectra are rectified 
to a mean flux value of unity in the 7000 to 7450\AA band (vertical lines).
\label{stars_flxmor}}
\end{figure}

A potentially important clue as to the nature of the blue/near-UV band discrepancy
is its variation with stellar parameters.  We have computed spherical LTE models
for four late-type giants for which we found reliable stellar parameters 
compiled in \citet{morossi_fmkb93} and spectrophotometric data in the catalogues
of \citet{pulk5} or \citet{burn}.  The \citet{morossi_fmkb93} list only 
contains stars of $[{\rm A}/{\rm H}]\approx 0$ 
with a quoted uncertainty of $\pm 0.25$, so they are more metal rich than 
Arcturus.  
The stars span a range in spectral type
from late G to mid K, but are all within four spectral sub-types of Arcturus,
and a range in luminosity class from sub-giant to bright giant.
\objectname{HR165} is almost an exact Arcturus analogue, but
has a higher metallicity than Arcturus.   
Details of the stars and spectrophotometric data sources are presented in 
Table 5, and the comparison of observed and theoretical spectra is shown in Fig.
\ref{stars_flxmor}.  Because the purpose of these comparisons is to investigate
the goodness of fit in the blue and near UV band with respect to that in the red
band, we arbitrarily scaled all spectra to produce an average flux level 
of unity in the 
7000 to 7400 \AA~ band.   

\paragraph{}
For HR165 (K3\,{\sc III}), \objectname{HR6869} (G8\,{\sc IV}) and \objectname{HR6705} 
(K5\,{\sc III}),
theoretical spectra that are force fit
to the observed spectra in the red provide a very close match to the observed
spectrum throughout the blue and near UV bands.  
This result
suggests that 
the blue and near UV band discrepancy is metallicity dependent in that 
mildly metal poor red giants like Arcturus generally exhibit the 
discrepancy, whereas solar metallicity red giants do not (Fig. \ref{stars_flxmor}).    
Such a metallicity dependent UV band flux discrepancy suggests NLTE effects as 
the cause, but, as can be seen from Fig. \ref{stars_flxall}, NLTE effects
make the discrepancy {\it worse}.  Note that spherical {\it NLTE} models may not
necessarily fit these solar metallicity stars as well as spherical LTE models do;
our conclusions are based on a {\it differential} assessment of the fit provided
by LTE models to mildly metal poor stars on one hand, and to solar metallicity stars 
on the other hand.

\subsection{UV opacity}

The prediction of too much flux in the near UV band of Arcturus was also found by 
\citet{short_l96} on the basis of $PP$ LTE modeling 
with the ATLAS9 code of \citep{kurucz92a} and SPECSYN LTE spectrum synthesis
code of \citet{kurucz_a81}.  
There has been much discussion of ``missing UV opacity'' in the models
of the Sun (see \citet{kurucz90}, \citet{kurucz92a} for example),  
and \citet{kurucz92a} achieved the first close fit to the {\it solar} UV flux
level at low resolution by incorporating the opacity of tens of millions of new 
theoretically predicted atomic lines. 
\citet{short_l96} suggested that the discrepancy may be due 
to the presence of an unknown source of 
{\it continuous absorption} opacity, $\kappa_{\rm Cnt}$, in the star that is 
unaccounted for in the atmospheric model.  
They doubled the amount of violet and near UV band continuous opacity in
their red giant model, {\it ad hoc}, and showed that the additional opacity
lowered the predicted violet $F_\lambda$ such
that it provided a much closer match to the observed $F_\lambda$ level. 
\citet{short_l96} also showed the addition of extra opacity in the
region below 4000\AA~ had no significant effect on the equilibrium structure
of the atmosphere because most of the flux is at longer $\lambda$ values.
A limitation of the \citet{short_l96} study was the use of $PP$ LTE models.
However, the more realistic models presented here over-predict
the blue and near UV flux by {\it even more} than the LTE $PP$ models.
{\it Therefore, the more realistic models in no way address the UV flux
problem for red giants, and do not obviate the need for increased 
opacity that was suggested by \citet{short_l96}.}

\section{Conclusions \label{sec_con}}

   We have presented atmospheric models and synthetic flux spectra
for the red giant star Arcturus that represent
two degrees of realism in the treatment of the atmospheric geometry and three 
degrees of realism in the treatment of the equilibrium state of the gas and 
the radiation field.  We have studied the ability of these models to fit
the overall absolute $F_\lambda(\lambda)$ level.  
We find that NLTE 
over-ionization of \ion{Fe}{1} substantially reduces the UV line opacity and allows
more flux to escape there. 
NLTE models predict approximately $50\%$ greater flux in the blue and near UV
band than LTE models.
Both LTE and NLTE models predict as much as $100$ to $150\%$ larger
flux than observed in the blue and near UV flux bands
This excess of model UV flux may indicate that there is {\it still}
substantial UV opacity missing from the models, despite the
greatly expanded line lists of \citet{kurucz92a}.  The extent of
the UV excess depends on whether
the Fe group elements are treated in NLTE.  
Also, we find that for the $[{\rm A}/{\rm Fe}]$ range 
-0.4 to -0.7 the effect of NLTE on the overall UV flux level {\it decreases} 
with decreasing metallicity. 

\paragraph{}

We calculated spherical
LTE models for late G and early to mid-K giants of {\it solar} metallicity
with reliable stellar parameters,
and found that they generally provided a close match to the observed
flux distributions for such stars from the near IR to the near UV band. 
Therefore, the blue/near-UV discrepancy is likely metallicity dependent
in the sense that it increases with decreasing metallicity.  Such a 
metallicity dependent flux discrepancy between observed and modeled 
flux distributions suggests that NLTE effects are the cause.  However, we
find that treating a large amount of the line opacity in self-consistent NLTE
{\it increases} the size of the discrepancy.   

\subsection{Future work}

  Given the results presented here, we are currently working on expanding our
NLTE$_{\rm Fe}$ calculations to span a range of stellar parameter space
among late G and early K giants.  Our goal is to map out in more detail the
metallicity dependency of the blue/near-UV flux discrepancy by applying NLTE models
with the highest level of realism to all such stars with observed global 
flux distributions.  The satisfactory resolution of this discrepancy 
is necessary for accurate modeling of the recently discovered XMP stars.

\acknowledgments

CIS gratefully acknowledges funding from Saint Mary's University and from the Charles E. Schmidt College 
of Science at Florida Atlantic University.  This
work was supported in part by  NSF grant AST-9720704, NASA ATP grant
NAG 5-8425 and LTSA grant NAG 5-3619 to the University of Georgia.
PHH was supported in part by the P\^ole Scientifique de Mod\'elisation
Num\'erique at ENS-Lyon. Some of the calculations presented in this
paper were performed at the San Diego Supercomputer Center (SDSC),
supported by the NSF, on the IBM pSeries 690 of the Norddeutscher
Verbund f\"ur Hoch- und H\"ochstleistungsrechnen (HLRN), and at the
National Energy Research Supercomputer Center (NERSC), supported by
the U.S. DOE. We thank all of these institutions for a generous
allocation of computer time.




\clearpage

\begin{deluxetable}{lrrrrrrr}
\footnotesize
\tablecaption{Species treated in Non-Local Thermodynamic Equilibrium (NLTE)
in the NLTE$_{\rm Light}$ and NLTE$_{\rm Fe}$ models.
Each ionization stage is labeled with the number of energy levels and bound-bound
transitions included in the statistical equilibrium rate equations.  Note that
this table shows only a sub-set of the total number of species that are 
currently treatable in statistical equilibrium by Phoenix. }
\tablecomments{Elements in bold face have been added since the last NLTE {\tt PHOENIX} modeling of red giant stars.
Those in italics have had their treatment improved since the last modeling.
}
\label{t1d}
\tablecolumns{8}
\tablewidth{0pt}
\tablehead{
\colhead{Element} & Model & \multicolumn{3}{c}{Ionization Stage} \\
\colhead{} & \colhead{} & \colhead{\ion{}{1}} & \colhead{\ion{}{2}} & \colhead{\ion{}{3}} }
\startdata
{\it H}   & NLTE$_{\rm Light}$, NLTE$_{\rm Fe}$     &  {\it 80/3160} &\nodata &\nodata  \\
He  & NLTE$_{\rm Light}$, NLTE$_{\rm Fe}$     &  19/37 & \nodata &\nodata  \\
Li  & NLTE$_{\rm Light}$, NLTE$_{\rm Fe}$     &  57/333 & 55/124 &\nodata  \\
C   & NLTE$_{\rm Light}$, NLTE$_{\rm Fe}$     &  228/1387 & \nodata & \nodata \\
N   & NLTE$_{\rm Light}$, NLTE$_{\rm Fe}$     &  252/2313 & \nodata & \nodata \\
O   & NLTE$_{\rm Light}$, NLTE$_{\rm Fe}$     &  36/66 & \nodata & \nodata \\
Ne  & NLTE$_{\rm Light}$, NLTE$_{\rm Fe}$     &  26/37 &\nodata &\nodata \\
{\bf Na}  & NLTE$_{\rm Light}$, NLTE$_{\rm Fe}$     &  {\bf 53/142} & 35/171 & \nodata \\
{\bf Mg}  & NLTE$_{\rm Light}$, NLTE$_{\rm Fe}$     &  {\bf 273/835} & 72/340 & \nodata \\
{\bf Al}  & NLTE$_{\rm Light}$, NLTE$_{\rm Fe}$     &  {\bf 111/250} & {\bf 188/1674} & \nodata \\
Si  & NLTE$_{\rm Light}$, NLTE$_{\rm Fe}$     &  329/1871 & 93/436 & \nodata \\
{\bf P}   & NLTE$_{\rm Light}$, NLTE$_{\rm Fe}$     &  {\bf 229/903} & {\bf 89/760} & \nodata \\
S   & NLTE$_{\rm Light}$, NLTE$_{\rm Fe}$     &  146/439 & 84/444 & \nodata \\
{\bf K}   & NLTE$_{\rm Light}$, NLTE$_{\rm Fe}$     &  {\bf 73/210} & {\bf 22/66} & \nodata\\
Ca  & NLTE$_{\rm Light}$, NLTE$_{\rm Fe}$     &  194/1029 & 87/455 & 150/1661 \\
Ti  & NLTE$_{\rm Fe}$     &  395/5279 & 204/2399 & \nodata \\
Fe  & NLTE$_{\rm Fe}$     &  494/6903 & 617/13675 & \nodata\\
\enddata
\end{deluxetable}

\clearpage

\begin{table}
\caption{Levels of modeling realism.}
\label{t2}
\begin{tabular}{lll}
\tableline
Degree of NLTE          & \multicolumn{2}{c}{Geometry}\\
                        & Plane-Parallel    & Spherical \\
\tableline
None                    & {\small LTE($PP$)} & {\small LTE} \\
Light metals only       & {\small NLTE$_{\rm Light}$($PP$)} & {\small NLTE$_{\rm Light}$} \\
Light metals \& Fe+Ti   & {\small NLTE$_{\rm Fe}$($PP$)} & {\small NLTE$_{\rm Fe}$} \\
\tableline
\end{tabular}
\end{table}

\clearpage

\begin{table}
\caption{Model grid.}
\label{t3}
\begin{tabular}{llll}
\tableline
Designation & $T_{\rm eff}$ & $\log g$ & $[{\rm A}/{\rm H}]$ \\
\tableline
T43G20M07\tablenotemark{a}   & 4300          & 2.0      & -0.7 \\ 
T43G15M07   & 4300          & 1.5      & -0.7 \\ 
T43G20M04   & 4300          & 2.0      & -0.4 \\
T43G15M04   & 4300          & 1.5      & -0.4 \\ 
T42G20M07   & 4200          & 2.0      & -0.7 \\ 
\tableline
\end{tabular}
\tablenotetext{a}{Canonical model}
\end{table}

\clearpage

\begin{table}
\caption{Pulkovo Spectrophotometric Catalogue stars that are Arcturus analogues.}
\label{t4}
\begin{tabular}{llll}
\tableline
HR number & Flamsteed & Spectral type \\  
\tableline
2260   &                    & K3 III \\
4232   &  $\nu$ Hya         & K2 III  \\
4630   &  $\epsilon$ Crv    & K2.5 IIIa  \\
6223   &  18 Dra            & K0 III \\ 
6859   &  $\delta$ Sgr      & K3 IIIa \\
\tableline
\end{tabular}
\end{table}

\clearpage

\begin{table}
\caption{Late-type giant stars of $[{\rm A}/{\rm H}]=0.0$ with reliable parameters \citep{morossi_fmkb93}.}
\label{t5}
\begin{tabular}{llllll}
\tableline
HR number & Flamsteed & Spectral type & $T_{\rm eff}$ & $\log g$ & Data source\\
\tableline
165  & $\delta$ And & K3\,{\sc III} & 4300 & 2.0 & \citet{pulk5}\\
6705 & $\gamma$ Dra & K5\,{\sc III} & 3900 & 1.5 & \citet{pulk5}\\
6869 & $\eta$ Ser & G8IV & 4900 & 3.0 & \citet{pulk5}\\
\tableline
\end{tabular}
\end{table}

\clearpage


\clearpage 

%

\clearpage 

\clearpage 

%

\clearpage 

\clearpage

\clearpage

%

\clearpage

\end{document}